\documentclass[twocolumn,showpacs,preprintnumbers,amsmath,amssymb]{revtex4}

\usepackage{graphicx}
\usepackage{dcolumn}
\usepackage{bm}

\begin{document}


\title{Experimental realization of large-alphabet quantum key distribution protocol using 
 orbital angular momentum entanglement}
\author{Shengmei Zhao$^{1,2}$}\email{zhaosm@njupt.edu.cn}
\author{Longyan Gong$^{1,3}$}\thanks{Corresponding author:LY Gong, lygong@njupt.edu.cn}
\author{Yongqiang Li$^{1}$, Hua Yang$^{1}$, Yubo Sheng$^1$, Xiaoliang Dong$^1$, Fei Cao$^{1}$}
\author{Baoyu Zheng$^1$}

\affiliation{$^{1}$Institute of Signal Processing and
Transmission, Nanjing University of Posts and
Telecommunications, Nanjing 210003, China \\
$^{2}$Key Lab of Broadband Wireless Communication and Sensor
Network Technology, Nanjing University of Posts and
Telecommunications, Ministry of Education, Nanjing 210003,
China\\
$^{3}$College of Science, Nanjing University of Posts and
Telecommunications, Nanjing 210003, China }
\date{today}
\begin{abstract} We experimentally demonstrate a quantum key distribution (QKD)
protocol using photon pairs entangled in orbit angular momentum
(OAM). In our protocol, Alice and Bob modulate their OAM states on 
each entangled pair with spatial light modulators (SLMs),
respectively. Alice uses a fixed phase hologram in her SLM, while
Bob designs $N$ different suitable phase holograms and uses them
to represent his $N$-based information in his SLM. With
coincidences, Alice can fully retrieve the key stream sent by Bob
without information reconciliation or privacy amplification.
We report the experiment results with $N=3$ and
the sector states with OAM eigenmodes $\left| \ell=1
\right\rangle$ and $\left| \ell=-1 \right\rangle$.  Our experiment shows that 
the coincidence rates are in relatively distinct value regions for 
the three different key elements.
Alice could recover fully Bob's keys by the protocol. 
Finally, we discuss
the security of the protocol both form the 
light way and against the general attacks.
\end{abstract}

\maketitle
\section{Introduction}\label{Sec1}
Quantum key distribution (QKD) allows two parties (typically
called Alice and Bob) to generate a secret key in the presence of
an eavesdropper (Eve). It performs the detection of any
eavesdropping, and the security is guaranteed by the fundamental
laws of quantum mechanics, such as, noncloning theorem and
Heisenberg uncertainty (see ref.\cite{SC09} and refs. therein).
QKD has attracted a great deal of research interest since BB84 QKD
protocol was reported \cite{BE84}.

Up to now, there have already been several typical QKD
protocols \cite{SC09,BE84,BE92,SC04,EK91,BE922,GR02,IN02}. For example, BB84 \cite{BE84}, B92 \cite{BE92}
and SARG04 \cite{SC04} protocols are nonorthogonal single-photon
protocols. E91 protocol \cite{EK91}, 
 and BBM protocol \cite{BE922}
are based on entangled photons. GG02 protocol \cite{GR02} and
distributed-phase-reference (DPR) protocol \cite{IN02} are
continuous-variable protocols. 
Currently, long-distance QKD over
$250$ kilometers experiments  have been reported \cite{ST09} over fiber, and secure 
transmission of a quantum key has been performed over 144 km in free space \cite{ursin2007}.
Meanwhile, some QKD systems have already been commercialized \cite{CO3}.
\par
Most of these QKD protocols rely on
either polarization or phase of faint laser pulse as information
carrier \cite{HA05}, the use of polarization encoding
limits the degree of a photon to encode information. 
Unlike the limited degree of freedom on polarization or phase
states, orbital angular momentum (OAM) states have attracted much
attention since there is an infinite OAM eigenstates in a
single-photon \cite{WA04, MT2007,Julio2008, LE09,LE10,DA11}. In 2007, 
Molina-Terriza \itshape et al.\upshape \ have 
reviewed the use of OAM of photons and discussed its potential for 
realizing high-dimensional quantum spaces \cite{MT2007}.  
With the help of  OAM entanglement, the group of Kwiat has beaten the 
channel capacity limit for linear photonic superdense coding \cite {Julio2008}. 
In 2010, Leach \itshape et al.\upshape \ demonstrated the quantum correlations 
in optical angle-orbital angular momentum variables. Their experimental results of 
angular EPR correlations have established that angular position and angular 
momentum are suitable variables for applications in current 
quantum information processing. They also experimentally demonstrated the violation of a 
Bell inequality in two and eleventh dimensional OAM state-space \cite{LE09,DA11}.
\par 
Apparently, the high capacity property of OAM states results in their applications in QKD protocol \cite{GR06,Gruneisen2008,Malik2012}. 
Recently, Gruneisen \itshape et al. \upshape\ proposed mutually unbiased 
bases (MUBs) with OAM for three dimensional quantum key distribution \cite{Gruneisen2008}. Malik 
\itshape et al. \upshape \ demonstrated an experimental implementation of a 
free-space 11-dimensional QKD system using OAM states \cite{Malik2012}. 
In these protocols, OAM states are used to design the
high-dimensional mutually unbiased bases. They are in principle the extension of BB84 protocol.
\par 
In this paper, we demonstrate a application for OAM entanglement 
in quantum key distribution.  We experimentally realize a large-alphabet 
QKD protocol using OAM entanglement. 
This QKD protocol is quite different from the traditional one because the 
protocol is dependent on the OAM entanglement, and the security is ensured 
by the property of entangled photon pairs and the special information modulation.
With the protocol, Alice and Bob can share keys without information reconciliation or privacy
amplification. Meanwhile, there is no classical channel in the protocol. 

This paper is organized as follows. In Sec. \ref{Sec2}, we detail
the encoded states, and then present our QKD protocol. In Sec. \ref{Sec3}, we give the
experimental results that certify our QKD protocol. In Sec.
\ref{Sec4}, we present the security analysis of the protocol. At
last, in Sec. \ref{Sec5}, we discuss the results and draw our
conclusions.

\section{Quantum key distribution protocol using OAM entanglement}\label{Sec2}
Our proposed QKD protocol is dependent on  the
the encoded states. We will first introduce
it in this section.

\subsection{Encoded states}
\begin{figure}[!htbp]
\begin{center}
 \includegraphics[width=2.5in]{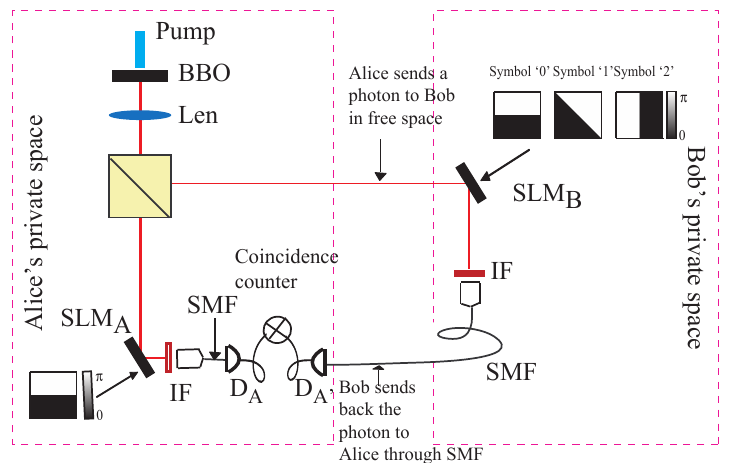}
\caption{(color online) Setup for the proposed QKD scheme.
}\label{Fig1}
\end{center}
\end{figure}
 
 The configuration considered for the protocol is shown in Fig.~\ref{Fig1}. 
The pump source is a Laguerre-Gaussian (LG) beam, which has zero
orbital angular momentum. As OAM is conserved, the two-photon
states generated from BBO crystal by SPDC process can be represented
by\cite{WA04,LE09,LE10,DA11,MA01}
\begin{equation}
\left| \Psi  \right\rangle  = \sum\limits_{\ell =  - \infty
}^{\ell = \infty } {\left| \phi \right\rangle _\ell}=
\sum\limits_{\ell =  - \infty }^{\ell = \infty } {c_\ell \left| +
\ell \right\rangle _A \otimes \left| { - \ell } \right\rangle _B
},
 \label{EQ1}
 \end{equation}
where subscripts $A$ and $B$ denote signal (Alice's) and idler
(Bob's) photons respectively, $|c_\ell|^2$ is the probability to
generate an entangled photon pairs (one photon in signal arm with
OAM $+\ell\hbar$ and the other photon in idler arm with OAM
$-\ell\hbar$), $\left| \ell \right\rangle$ is the OAM eigenmode
with mode number $\ell$, and $\hbar$ is the Plank constant divided
by $2\pi$. The OAM state of beam could then be
changed by a phase filter, which is implemented by
computer-controlled SLMs \cite{LE05}. Here, the SLMs act as
reconfigurable holograms. The modulated photons is collected via a
single-mode fiber (SMF) which is fed to a single photon detector.
As only the OAM eigenmodes $\left| \ell=0 \right\rangle$ couples
into the fiber (selected only for $\left| \ell=0 \right\rangle$),
a count in the detector indicates a detection of the state in
which the SLM was prepared for \cite{DA11}, where $\ell$ is the
azimuthal index for OAM states of photons\cite{GR06}. At last, the
outputs from the detectors are fed to a coincidence counter. 
Experiments have justified that the generated two
photons are in entanglement \cite{WA04,LE09,LE10,MA01,DA11}. Using
the modified Pointcar\'{e} sphere in an analogous fashion to
polarization states, we use the sector state to encode symbol sequence in our QKD protocol. 
\par
A sector sector state is defined as an equally
weighted superposition of $\left| \ell \right\rangle$ and $\left|
-\ell \right\rangle$ with an arbitrary relative phase, which is
represented by a point along the equator \cite{LE09}
\begin{equation}
\left|\theta_\ell \right\rangle=\frac{\sqrt{2}}{2}
(e^{i\ell\theta}\left|+\ell\right\rangle+e^{-i\ell\theta}\left|-\ell\right\rangle),
 \label{EQ2}
 \end{equation}
 where $\theta$ relates to the orientation of the sector state in the
 modified Pointcar\'{e} sphere. Actually, it is the angle between the 
 sector state and the superposition state $\frac{\sqrt{2}}{2}
(\left|+\ell\right\rangle+\left|-\ell\right\rangle)$ on the equator.

For an entangled photon pairs, the coincidence of one photon in
sector state $\left|\theta_{A,\ell} \right\rangle$ and the other in state
 $\left| \theta_{B,\ell} \right\rangle$ is \cite{LE09}
\begin{equation}
\begin{split}
C(\theta_{A,\ell},\theta_{B,\ell})= & |<\theta_{A,\ell}|<\theta_{B,\ell}|\Psi>|^2 \\
                                    & \propto \cos^2{[\ell(\theta_A-\theta_B)]}. \label{EQ3}
\end{split}
\end{equation}
The high-visibility sinusoidal fringes of this joint probability could be used 
for the symbol retrieving. For example, if  a  $\left|\theta_{A,\ell} \right\rangle$ is fixed 
for one photon at Alice's side, and some discrete $\left|\theta_{B,\ell} \right\rangle$ 
with different angles are prepared for the other photon at Bob's side, 
the coincidence will be different for the different angles. 
At the same time, if the different angles here are the representation 
of different information at Bob's side, one could judge the different 
inforamtion from the different coincidence rates at Alice's side. 
We use these sector states to encode key sequences at Bob's side and recover 
the symbol information from the coincidence rate at Alice's side in our proposed QKD protocol.

\subsection{Quantum key distribution protocol}

As shown in the sketch of the setup in Fig.~\ref{Fig1}, the
regions in the left box and the right box are Alice's and Bob's
private places in our protocol, respectively. 
Based on the configuration and the sector states
shown in Eq.(\ref{EQ2}), the proposed QKD protocol is designed as
follows.

(1) With BBO crystal, Alice generates entangled photon pairs,
named signal photons and idler photons. Alice keeps the signal
photons for herself and sends the idler photons in the free-space
channel to Bob.

(2) Alice selects a sector state $\left|\theta_{A,\ell}
\right\rangle$ for the signal photons. Here, the sector state is 
implemented by a computer-controlled
$SLM_A$ with a special hologram in combination with SMF on the
signal photons.

(3) At the same time, Bob will encode his symbol sequence with his sector states.
He also selects the sector states using $SLM_B$
with different phase holograms in combination with SMF on the
idler photons. The corresponding sector states are
$\left|\theta^0_{B,\ell} \right\rangle, \left|\theta^1_{B,\ell}
\right\rangle, ..., \left|\theta^{N-1}_{B,\ell} \right\rangle$,
which are codes for key elements``$0$",``$1$",...,``$N-1$",
respectively. For a stream of key elements, Bob successively
chooses the corresponding phase hologram  and modulate them in his
SLM. For each key element, the corresponding modulation continues
for a regular time interval $\tau$ for enough coincidence
counting. Of course, if the detectors have high enough detection efficiency, the regular 
time interval is not needed.

(4) Alice and Bob transmit the modulated photons to the detectors
$D_A$ and $D_A'$ at Alice's private place through a single mode
fiber, respectively.

(5) Alice performs a coincidence measurement at her site. According to
the different coincidence rates, Alice can retrieve the key sequences prepared by Bob, so that they 
could share the same keys. Note that the suitable modulations for Bob in the
\emph{3rd} step, mean that in the \emph{5th }step, the
corresponding coincidences expressed in Eq.(\ref{EQ3}) are in
distinct value regions.
\\

On the other hand, in the above $1$---$5$ steps, Alice and Bob
replace each other, then Bob can recover keys sent by Alice. In
other words, if there are two such setups, Alice and Bob can send
keys to each other.

\section{Experimental Results}\label{Sec3}
\begin{figure}[!htbp]
\begin{center}
 \includegraphics[width=2.5in]{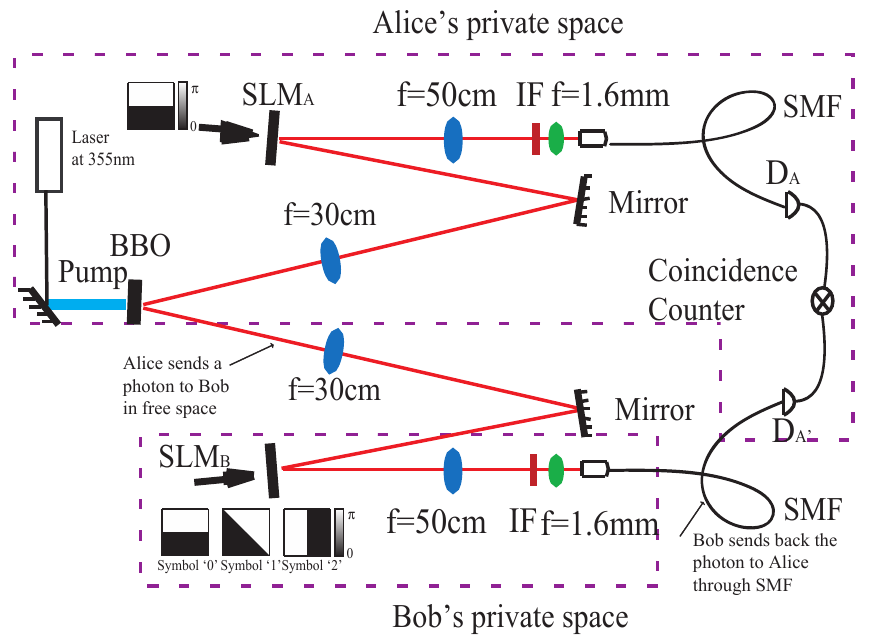}
\caption{(color online) Setup for the proposed QKD scheme.
BBO:$BaB_2O_4$ nonlinear crystals; $SLM_A$:spatial light modulator
for Signal photons; $SLM_B$:spatial light modulator for idler
photons; SMF:single mode fiber; $D_{A}$:detector for signal
photons; $D_{A'}$:detector for idler photons. In our protocol, the
upper box and the lower box are Alice's and Bob's private places,
respectively.}\label{Fig2}
\end{center}
\end{figure}

In order to testify this protocol by experiment, we setup an 
experimental system as Fig.~\ref{Fig2}, where a quasi
continuous-wave, mode-locked $(100 MHz)\ 355 nm$ laser is selected
as the pump source. It is focused into a nonlinear crystal
[$\beta$-barium borate (BBO)], which can generate OAM entangled
photons through a frequency-degenerate type-I spontaneous
parametric down-conversion (SPDC) process. 
The spatial light modulator are the product from Hamamatsu Photonics. 
 The modulated photons is collected via a
single-mode fiber (SMF) with $5 um$. The single photon detectors 
are the detection modules from Perkin Elmer. The interference filters selected
are at $710nm$ with $10nm$ bandwidth (Thorslab product). In addition, some lens are used to 
illustrate the beam path is long enough in free space. Finally, 
The outputs from the detector modules are fed to 
a self-made coincidence counter. In order to implement the experiment, we 
let the two side (Alice and Bob) be symmetric.   
The detail of the experimental setup is shown as Fig.~\ref{Fig2}.

To verify the above protocol, we give experimental results at
$N=3$ and the sector states in Eq.(\ref{EQ2}) with $\ell=1$. In
the \emph{3rd} step of the protocol, we set the regular time
interval $\tau=1000ms$ , which guarantees to produce no less than
$500$ entangled photon pairs in the time interval. Henceforth, we
omit $\ell$ for simplicity. We choose
$\left|\theta_A=\frac{\pi}{2} \right\rangle$, $\left|\theta^0_B=0
\right\rangle$, $\left|\theta^1_B=\frac{\pi}{4} \right\rangle$,
$\left|\theta^2_B=\frac{\pi}{2} \right\rangle$. From
Eq.(\ref{EQ3}), $C(\theta_A=\frac{\pi}{2},\theta^0_B=0)$ is
relatively small,
$C(\theta_A=\frac{\pi}{2},\theta^1_B=\frac{\pi}{4})$ is middle,
and $C(\theta_A=\frac{\pi}{2},\theta^2_B=\frac{\pi}{2})$ is
relatively large.

\begin{figure}[!htbp]
\begin{center}
 \includegraphics[width=2.5in]{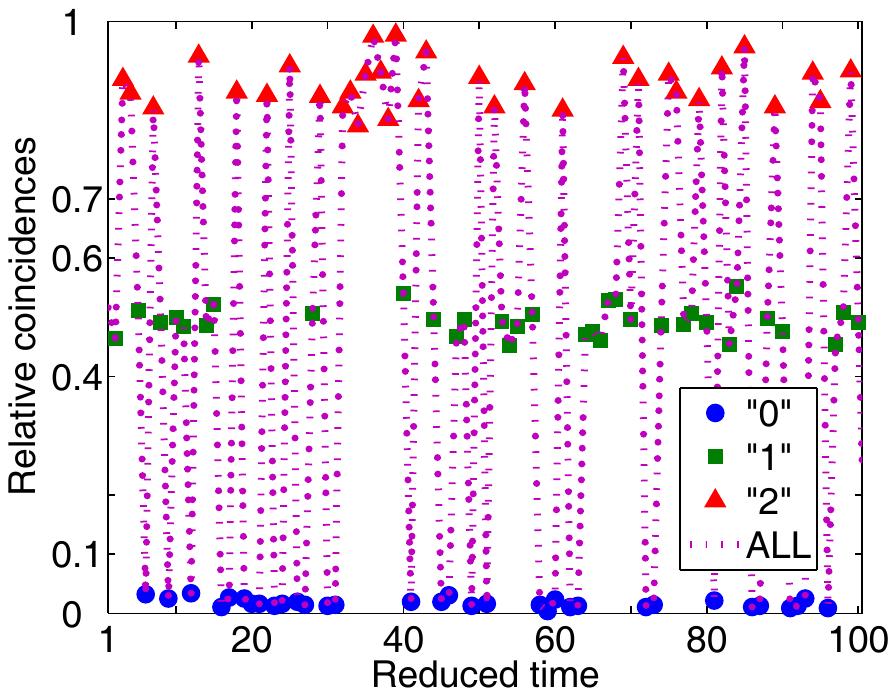}
\caption{(color online) The relative coincidences vary with the
reduced time for key elements``$0$",``$1$" and ``$2$",
respectively.} \label{Fig3}
\end{center}
\end{figure}

In Fig.\ref{Fig3}, we show the relative coincidences versus the
reduced time , which are for $100$ successive keys from a key
stream with length $10000$. Here the relative coincidence is the
ratio of photon coincidence counting during $\tau$ to the maximal
counting $561$ in our experiment, and the reduced time is  the
real time divided by $\tau$. We find that the relative
coincidences are near \emph{zero} for the key element ``0'',  near
\emph{one} for the key element ``2'', and around \emph{one-half}
for the key element ``1'', which agrees with the calculated
results from Eq.(\ref{EQ3}).

\begin{figure}[!htbp]
\begin{center}
 \includegraphics[width=2.5in]{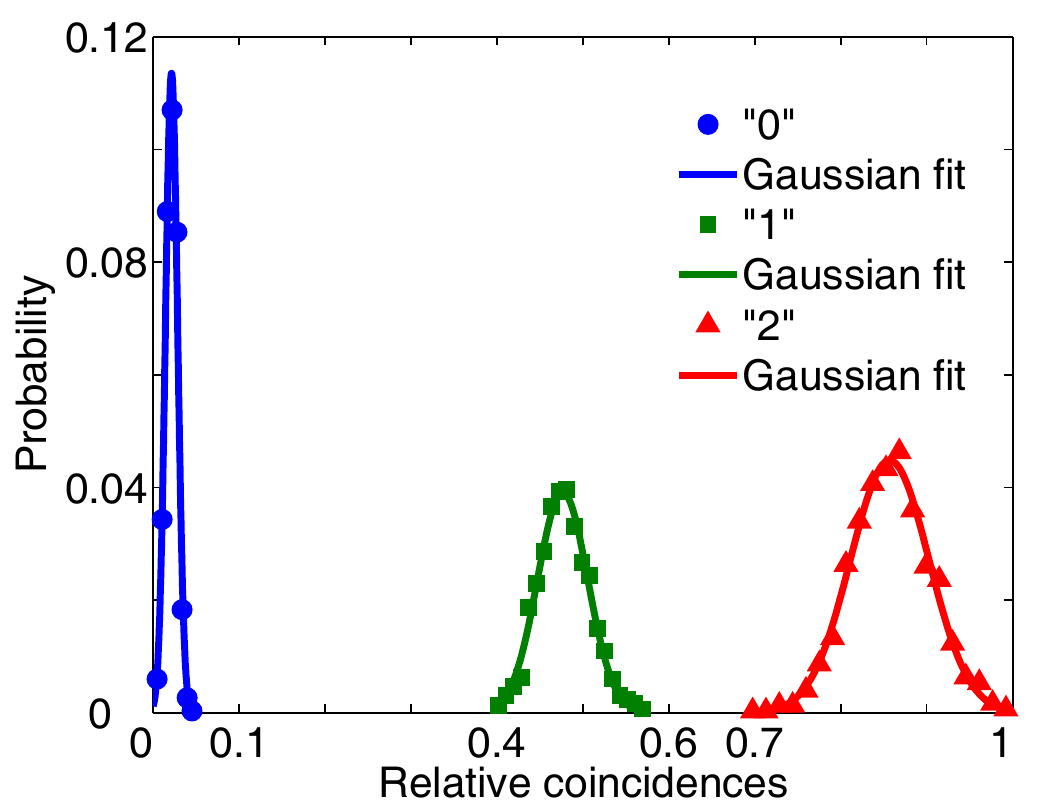}
\caption{(color online) The probability of relative coincidences
for key elements``$0$",``$1$" and ``$2$", respectively. The curves
are corresponding Gaussian fits. }\label{Fig4}
\end{center}
\end{figure}

\begin{table*}
\caption{the partial experiment results and corresponding keys at
$\ell=1$}\centering
\begin{tabular}
{|p{3.4cm}|p{0.65cm}|p{0.65cm}|p{0.65cm}|p{0.65cm}|p{0.65cm}|p{0.65cm}
|p{0.65cm}|p{0.65cm}
|p{0.65cm}|p{0.65cm}|p{0.65cm}|p{0.65cm}|p{0.65cm}|} 
\hline Bob's key   & $0$    & $1$    & $1$    & $2$ & $2$ & $1$ &
$0$ & $2$ & $0$  & $1$  & $0$ & $2$ & $1$\\ \hline $\theta_B$ &
$0$ & $\frac{\pi}{4}$&$\frac{\pi}{4}$&
$\frac{\pi}{2}$&$\frac{\pi}{2}$&$\frac{\pi}{4}$&
$0$&$\frac{\pi}{2}$& $0$&$\frac{\pi}{4}$&$0$&$\frac{\pi}{2}$&
$\frac{\pi}{4}$\\
\hline $\theta_A$& $\frac{\pi}{2}$ &$\frac{\pi}{2}$
&$\frac{\pi}{2}$ &$\frac{\pi}{2}$ &$\frac{\pi}{2}$&$\frac{\pi}{2}$
&$\frac{\pi}{2}$&$\frac{\pi}{2}$ &$\frac{\pi}{2}$&$\frac{\pi}{2}$
&$\frac{\pi}{2}$&$\frac{\pi}{2}$ &$\frac{\pi}{2}$ \\ \hline
coincidence  &$4$ &$291$ &$260$ &$506$&$493$
&$287$&$18$&$479$&$14$&$280$&$19$&$528$ &$293$\\ \hline Relative
coincidence &$0.007$&$0.519$&$0.464$&
$0.902$&$0.879$&$0.512$&$0.032$&$0.854$&$0.025$&$0.499$&$0.034$&
$0.941$&$0.522$\\ \hline key recovered by Alice
&$0$&$1$&$1$&$2$&$2$&$1$&$0$&$2$&$0$&$1$&$0$&$2$&$1$\\
\hline
\end{tabular}
\label{table2}
\end{table*}

In Fig.\ref{Fig4}, we give the probability of relative
coincidences for key elements ``$0$", ``$1$" and ``$2$", as a
statistic result from an experiment containing $10000$ keys. It
shows that they all obey Gaussian distribution due to the Gaussian
noise of light sources. For key elements ``$0$",``$1$" and ``$2$",
the values of relative coincidences are in the ranges $[0,0.05],
[0.40,0.57]$ and $[0.70,1.0]$, respectively.  There are large
enough gaps between arbitrary two nearest ranges. In other words,
the coincidence rates are in distinct value regions for different
key elements. Therefore, Alice can judge the key element sent by
Bob according to which region the value of the coincidence rate
belongs to, i.e., Alice can recover Bob's keys. To test this, we
let Bob send another $10000$ keys in experiments. We find that
Alice can fully recover Bob's keys with the decision criteria. The
corresponding partial experimental results are illustrated in
Table I. 

\section{SECURITY ANALYSIS}\label{Sec4}
Next, we will present our argument about the security of our QKD
protocol from two aspects. One aspect is the security at the light
way, and the other is the security against typical attacks.

\subsection{Security at the light way}
In the protocol, Alice generates the signal and idler photons from
BBO crystal by SPDC. Then, Alice modulates a fixed phase on the
photons in order to select special OAM states from the
superposition of all OAM states from $\ell=-\infty$ to
$\ell=\infty$. The modulated photons travel in SMF. Last, Alice
take coincident measurement and extract the key from the
measurements. As shown in the left box in Fig.\ref{Fig1}, all
these happen in Alice's site. During all the procedures, Eve can
not access the signal photons. Therefore, the signal photons are
safe.

As shown in the right box in Fig.\ref{Fig1}, Bob modulate the
idler photons in his site. Eve can not access the idler photons
during Bob's modulation. They are safe at the above procedure.

There are two chances for Eve to access the idler photons when
idler photons travel in public quantum channels. As displayed in
Fig.\ref{Fig1}, one chance is when the idler photons are
transmitted from Alice to Bob in free spaces after Alice generates
the OAM entangled photons. The other is when the idler photons are
transmitted from Bob's private place to Alice's site in SMF after
Bob modulates his information. For simplicity, we call the two
chances the ATB chance and the BTA change, respectively. For the
ATB chance, at that time, the state carried by idler photons is a
superposition of all OAM states and does not have Bob's
information. Furthermore, the idler photons are entangled with the
signal photons. Hence, even Eve accesses these photons, she can
not steal any Bob's information from the idler photon. Actually,
Alice could detect this action since she keep
the other entangled photon. 
For the BTA chance, there is only
$\left|\ell=0\right\rangle$ mode in the idler photons when they
travel in SMF, and $\left|\ell=0\right\rangle$ mode has no key
information in Bob's encoding scheme. As entangled photon pairs
are used in our protocol, Bob's information can be recovered only
using coincidence measurement between idler and signal photons.
However, in our protocol, Eve has no change to access signal
photons , therefore, she also can not get any Bob's key information
at the procedure.

From all the above, we find that Eve can not get Bob's key
information for all the light way. Furthermore, the photon losses
are unavoidable when photons travel in light ways nearly for all
systems. Here Alice (Bob) recovers a key sent by Bob (Alice) using
coincidence measurements. The corresponding coincidence counting
rate is a statistical result, therefore, the value of key can also
be recovered even at the condition that small fraction of photons
loss.

\subsection{Security against typical attacks}
During QKD, there are many eavesdropping strategies for Eve to get
key information. The typical attacks are the intercept-resend (IR)
attack, the man-in-the-middle (MIRM) attack, and the
photon-number-splitting (PNS) attack, \emph{et. al}.  Eve can
perform these attacks in the ATB chance or the BTA chance that
mentioned above. In the following, we analysis them respectively.

(i)The IR attack: Eve measures out every signal emitted by Alice
and prepares a new one, depending on the result obtained, that is
given to Bob. As described above, in the ATB and the BTA chances,
there is no difference between these idler photons. Eve can
intercept and measure them. However, there is no meaning for these
measurement results. Therefore, Eve can not get any key
information. As entangled photon pairs are used, the photon resent
by Eve will not satisfied with Eq.(\ref{EQ3}), so Alice can detect
the eavesdropper.

(ii)The MIRM attack: Eve pretends to be Bob to Alice and
simultaneously pretends to be Alice to Bob.  QKD is vulnerable to
this attack when used without authentication.  In our protocol,
Alice (Bob) can fully recover Bob's (Alice's) keys. They can also
send deterministic keys to each other. If Alice and Bob have an
initial shared secret, they can use these deterministic keys to
authenticate each other.

(iii)The PNS attack: Weak pulses may contain more than one photon
and Eve can simply keep some of the photons while letting the
others go to Bob. In our protocol, many photons in a regular time
interval $\tau$ generate a qudit, which is equivalent to the
many-photon case. Eve can keep some of the photons while let the
others go to Bob during one $\tau$. As entangled photon pairs and
sector states are used, Eve can't get key information carried by
photons for she has no chance to perform coincidence measurements.

\section{Discussions and conclusions}\label{Sec5}

In the paper, we have experimentally demonstrated the proposed large-alphabet QKD protocol 
with the entangled-photon in Bob's side carrying trits of information. 
Theoretically, Bob uses OAM as information
carrier, so that he can encode more than one bits on his key sequences
every time. We have analyzed the security of the protocol both from 
the light way and the typical attacks. The results show that the security of 
the protocol is ensured by
the property of the entangled photons and the special fashion of
the information modulation. 
The experimental results also show that Alice could recover fully Bob's information, 
so that Alice and Bob can share keys without information reconciliation or privacy
amplification.

\par
In principle, our QKD protocol have some advantages over the traditional QKD
protocols. First, we use OAM states as the information carrier, 
there is in principle no limit to how many bits encoded in
each entangled photon pairs, which results in a potentially 
very large-alphabet size. Second, we utilize the entangled photons and 
the special fashion of the information modulation in the protocol, and recover 
the key sequences by coincidence counting. Since one photon is 
always kept in Alice's private space, Eve has no way to eavesdrop the key sequences 
during the protocol steps. 
All these assure the security of our proposed protocol. 
In addition, in the protocol, Alice and Bob do not need any classical
communications for the key recovery, which makes the protocol efficient and simple.

In practice, the bits encoded in
each entangled photon pairs is determined by
the light source properties, noise intensity and the photon detection to
distinguish the corresponding coincidences expressed in
Eq.(\ref{EQ3}). Experimentally, we could recover four-based keys with 
the distinct coincidences if we set the 
regular time interval $\tau=200ms$ so that more than $100$
entangled photon pairs could be produced during the time interval.  
Hence, we got $5$ four-based keys per second with $10$ bits per
second (bps) at this case. 
Currently, the coincidence rate is in the order $10^2$ to
$10^4$ counts per second using different entangled photon
sources, and the corresponding key generation rates are in the order
$10$ to $10^3$ bps with our QKD.  This is slower than QKD technologies based on weak coherent pulses.
But, with the development of
techniques about the photon detection and the generation of
entangled photon pairs, the key generation rate will be improved greatly.
\par
Generally, the practical implementing of QKD based on OAM 
states will confront two big problems, such as, decoherence by 
atmospheric turbulence and photon losses in quantum channel. 
Theoretically and experimentally, these two problems have been attracted a lot of attention \cite{Paterson2005, Pors2011}, and some solutions have been proposed to overcome these problems \cite{ZH12,Gisin2010}.

\par
Finally, as the keys can be recovered completely, the protocol can also be
used to quantum authentication, quantum direct communication \cite{BO02,DE03,LE06} and
so on. 
\par

\section{Acknowledgment}
The authors thank Optics group, School of Physics and Astronomy,
Glasgow University for hosting S.M. as a visitor. S.M. thanks Dr.
Leach for helping the setup of the experimental platform. S.M. is
partially supported by UNSRF (No.11KJA510002), PAPD of
JSHEI(No.NJ210002), and ORF of KLBWC\&SNT(MOE). L.Y. is supported
by NSFC (No.10904074).


\begin{thebibliography}{99}
\bibitem{SC09} V. Scarani, N. J. Cerf, M. Du\v{s}ek, N. L\"{u}tkenhaus,
and M. Peev, Rev. Mod. Phys. \textbf{81}, 1301 (2009).
\bibitem{BE84} C. H. Bennett and G. Brassard, 1984, \emph{Proceedings IEEE
International Conference on Computers, Systems and Signal
Processing, Bangalore, India }(IEEE, New York), p. 175.
\bibitem{BE92} C. H. Bennett, Phys. Rev. Lett. \textbf{68}, 3121 (1992).
\bibitem{SC04} V. Scarani, A. Ac\'{i}n, G. Ribordy, and N. Gisin, Phys.
Rev. Lett. \textbf{92}, 057901 (2004).
\bibitem{EK91} A. K. Ekert, Phys. Rev. Lett. \textbf{67}, 661 (1991).
\bibitem{BE922} C. H. Bennett, G. Brassard, and N. D. Mermin, Phys. Rev. Lett.
 \textbf{68}, 557 (1992).
\bibitem{GR02} F. Grosshans and P. Grangier, Phys. Rev. Lett.
\textbf{88}, 057902 (2002).
\bibitem{IN02} K. Inoue, E. Waks, and
Y. Yamamoto, Phys. Rev. Lett. \textbf{89}, 037902 (2002).
\bibitem{ST09} D. Stucki, N. Walenta, F. Vannel, R. T. Thew, N. Gisin, H.
Zbinden, S. Gray, C.R. Towery, and S. Ten, New J. Phys.
\textbf{11}, 075003 (2009).\bibitem{ursin2007} R.Ursin, F.Tiefenbacher, T.Schmitt-Manderback, H.Weier, 
T.Scheidl, M.Lindenthal, B.Blauensteiner, T.Jennewein, J.Perdigues, P.Trojek, 
B.\"{O}mer, M.F\"{u}rst, M.Meyenburg, J.Rarity, Z.Sodnik, C.Barbieri, H.Weinfurter, and A. Zeilinger, 
Nature Physics \textbf{3}, 481-486 (2007).
\bibitem{CO3}idQuantique, Geneva(Switzerland) (www.idquantique.com);
MagiQ Technologies, Inc., New York(www.magiqtech. com); and
Smartquantum, Lannion (France) (www. smartquantum.com).
\bibitem{HA05} Z. F. Han, X. F. Mo, Y. Z. Gui and G. C. Guo, Appl.
Phys. Lett. \textbf{86}, 221103 (2005).
\bibitem{WA04}S. P. Walborn, A. N. de Oliveira, R. S. Thebaldi, and C. H.
Monken, Phys. Rev. A \textbf{69}, 023811(2004). 
\bibitem{MT2007} G. Molina-Terriza, J. P. Torres, and L. Torner, Nature Physics,
\textbf{3}, 305-310 (2007).
\bibitem{Julio2008} J. T. Barreiro, T. C. Wei, and P. G. Kwiat, Nature Physics
\textbf{4}, 282-286 (2008). 
\bibitem{LE10}J. Leach, B. Jack, J. Romero, A. K. Jha, A. M. Yao, S.
Franke-Arnold, D. G. Ireland, R. W. Boyd, S. M. Barnett and M. J.
Padgett, Science \textbf{329}, 662 (2010).
\bibitem{LE09}J. Leach,
B. Jack, J. Romero, M. Ritsch-Marte, R. W. Boyd, A. K. Jha, S. M.
Barnett, S. Franke-Arnold, and M. J. Padgett, Opt. Exp.
\textbf{17}, 8287 (2009). 
\bibitem{DA11}A. C. Dada, J. Leach, G. S. Buller, M. J.
Padgett and E. Andersson, Nature Physics \textbf{7}, 677 (2011).
\bibitem{GR06}S. Gr\"{o}lacher, T. Jennewein, A. Vaziri, G. Weihs
and A. Zeilinger, New J. Phys. \textbf{8}, 75 (2006).
\bibitem{Gruneisen2008} M. T. Gruneisen, W. A. Miller, R. C. Dymale 
and A. M. Sweiti, Appl. Optics \textbf{47}(4), A 32 (2008).
\bibitem{Malik2012} M. Malik, M. O'Sullivan, B. Rodenburg, M. Mirhosseini, 
J. Leach, M. P. J. Lavery, M. J. Padgett and R. W. Boyd, arXiv:1204.5/81v1 (2012)
\bibitem{MA01}A. Mair, A. Vaziri, G. Weihs, and A. Zeilinger, Nature (London)
\textbf{442}, 313 (2001).
\bibitem{LE05}J. Leach, M. R. Dennis,
J. Courtial and M. J. Padgett, New J. Phys. \textbf{7}, 55 (2005).
\bibitem{Paterson2005} C. Paterson, Phys. Rew. Lett. \textbf{94}, 153901 (2005).
\bibitem{Pors2011} Bart-Jan Pors, C. H. Monken, Eric R. Eliel, and  J. P. Woerdman, 
Opt. Exp. \textbf{19}, 6671, 2011. 

\bibitem{ZH12}S. M. Zhao, J. Leach, L. Y. Gong, J. Ding, and B. Y. Zheng,
Opt. Exp. \textbf{20}, 452 (2012).

\bibitem{Gisin2010} N. Gisin, S. Pironio, and N. Sangouard, 
Phys. Rew. Lett. \textbf{105}, 070501 (2010).

\bibitem{BO02}K. Bostr\"{o}m and T. Felbinger, Phys.Rev. Lett.\textbf{89},187902 (2002).
\bibitem{DE03}F. G. Deng, G. L. Long, and X. S Liu, Phys. Rev. A \textbf{68},
042317 (2003).
\bibitem{LE06}H. Lee, J. Lim and H. J. Yang, Phys. Rev. A. \textbf{73}, 042305 (2006).

\end{thebibliography}
\end{document}